# Low-Profile Dual Band-pass Frequency Selective Surface with independent bands of operation

Komlan Payne[1]

[1]Department of Biomedical Engineering, State University of New York at Buffalo, Buffalo, NY 14260 USA
**Correspondence author, komlanpa@buffalo.edu**

*Abstract*— **A low-profile, modified Complementary Frequency Selective Surface (CFSS) with dual band-pass characteristic is presented. This technique adds independent control of the operation bands, which was a limitation from previous FSSs design based on CFSS concept. The FSS structure utilizes resonant elements and interaction between the layers to deliverer the first operating bandpass response at a low frequency. It is demonstrated that the first order response can achieved miniaturized periodic dimensions of about $\lambda_l/15$ and ultra-thin overall thickness compressed down to $\lambda_l/200$, where $\lambda_l$ is the free space wavelength at the lowest band of operation. An equivalent circuit model (ECM) is presented to explain the working principles of the proposed FSS. This investigation is further supported by parametric studies based on full wave simulation showing how this simple approach allows decoupling of the two bands of operation. A prototype, simultaneously operating at S and C bands is fabricated and tested in a free space environment for validation purpose.**

## I. INTRODUCTION

Frequency selective surfaces (FSS) are spatial filters composed of periodic structures arranged in one or two dimensional lattice [1]. It is constructed of arbitrary geometry of conductive patterns using single or stacked layers backed by dielectric materials. Some of the applications include radomes, dichroic surfaces for reflectors and subreflectors [2], low radar cross section and electromagnetic interference reduction in wireless indoor environment [3]. FSS are also integrated with active components and employed in missiles attack and electronic protection systems against high power microwave energy [4].

FSS with miniaturized elements (where the dimension of a unit cell are much smaller than the wavelength) provides stable filtering response for waves incident from oblique angles, thus are being extensively researched. Miniaturized element frequency selective surface (MEFSS) was proposed by cascading non-resonant periodic patches and wire grid layers [5]. An alternative design technique using cascaded miniaturized resonators coupled through aperture interlayer has also been reported recently [6], [7]. Aside the advantage of MEFSS, the ability of FSS to simultaneously operate in multiple bands has found interests in wireless and satellite communication link. Compact multiband devices are alternative solutions in systems where size is limited and multiservice is required. Several techniques including: fractal element [8], multi-resonant unit cells, cascading of single layer and complementary FSS achieve a multiband operation. An FSS consisting of a meandered wire loop on the top metal layer and its complementary pattern at the bottom layer but translated with an offset, demonstrates a dual-band operation [9]. A triple band FSS operating in C-/Ku-/Ka-bands is also presented in [10] consisting of periodic array of double metallic square loops and its complementary aperture loops. However, aforementioned FSS, not only suffer from a narrow passband (less than 10%) but also from either a very closed or significant band separation between the operation bands which cannot be controlled.

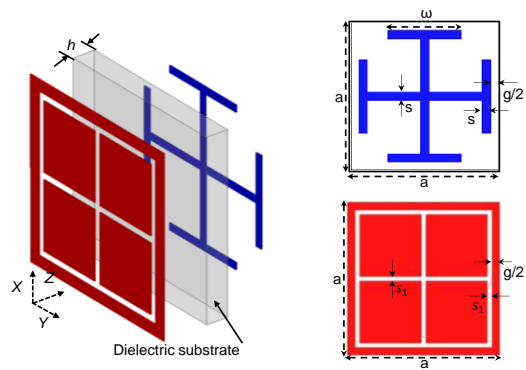

Fig. 1. Topology of the proposed first order FSS composed of subwavelength resonators element separated by a super thin dielectric material along with the detailed geometry of the unit cell of each layer.

In this communication, we demonstrate a low-profile dual-band FSS with higher degree of design freedom used to independently control the two poles and zeros of its transfer function. The design has been briefly introduced in [11] to demonstrate dual band operation and miniaturization feature. This present paper is an expansion of [11] with a comprehensive analysis to demonstrate the design capability to control the spectral gap between the bands. Also, in order to improve the filter selectivity at the upper band, a simple low profile structural design is realized to achieve higher order response. The specification of the design and model analysis are illustrated in section II with parametrization studies detailed in section III. In section IV, a first order FSS operating at S and C bands with center frequencies $f_l = 2.4$ GHz and $f_u = 5.8$ GHz respectively, is designed based on the outcomes obtained in section III. The prototype is then fabricated and its performance is experimentally validated. The dual-band FSS with higher order bandpass response and independent bands of operation is also presented in section V showing its versatile implementation. Finally, a summary of the present work is addressed in section VI.

## II. Design Specification and Model Analysis

The topology of the proposed first order dual-band FSS is shown in Fig. 1. The dielectric slab is sandwiched in between a bandpass and a bandstop layer. The design concept is adapted from the spatial filter introduced in [13], where a dual bandpass FSS was realized using complementary frequency selective surface (a bandpass layer and its complementary bandstop layer in a cascade connection separated by a thin dielectric slab). This design that consists of unit elements of resonant dimensions, however, does not allow flexible control of the passband or transmission zero locations as the constituting $L$ and $C$ cannot be independently controlled. The reason behind this limitation is that both layers are counterparts to each other in the way that if they are arranged together on top of each other, a complete conducting screen is obtained [1]. To remedy this problem, a bandpass and a bandstop layer composed of miniaturized unit elements that allow independent tuning of $L$ and $C$ is proposed. As shown in Fig. 1, the first layer, a 2D periodic array of cross slot dipole bounded by a square loop wire slot demonstrates a bandpass characteristic, whereas the second layer, an array of JC structure exhibits a band stop response. The top view of the unit cell of each filter layers is also denoted in Fig. 1. The working principle is similar to that of [13] but with better control of constituting reactance values as will be further discussed in Section III. The ECM for normal angle of incidence of a plane EM wave is shown in Fig. 2. The structure is modeled as a parallel $L_p$, $C_p$ tank with a parasitic inductor $L$ in parallel with a serie $L_s$, $C_s$ resonator separated by a short transmission line. The substrate with thickness $h$ and dielectric constant $\varepsilon_r$ is modeled by the short transmission line with length $h$ and characteristic impedance $Z_d = Z_o/\sqrt{\varepsilon_r}$. As for the upper and lower semi-infinite spaces on each side of the structure, they are also modeled by semi-infinite transmission line with characteristic impedance $Z_0 = 377\Omega$. Since $s_1 \ll a$, and the electrical length of the transmission line $\theta_d$ is very small at both bands, the parasitic inductor $L$ and the effect of the

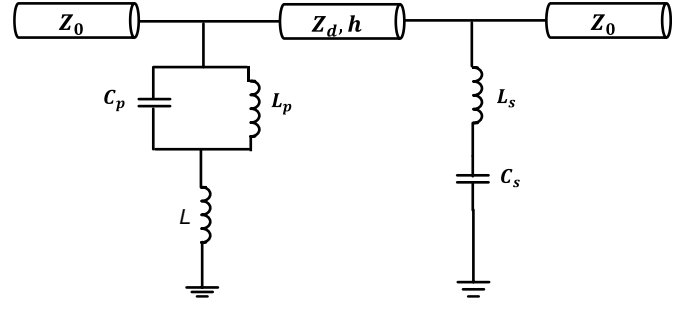

Fig. 2. Equivalent circuit model of the first order MEFSS proposed consisting of a tank resonator in parallel with a notch resonator separated with a short transmission line.

substrate can be ignored. Thus, a simple surface impedance of the design can be calculated from the ECM:

$$Z_s = \frac{j\omega L_p (1 - \omega^2 L_s\, C_s)}{1 - \omega^2 (L_p\, C_p + L_s\, C_s + L_p\, C_s) + \omega^4 L_p\, C_p\, L_s\, C_s} \quad (1)$$

It is clear that the impedance of the FSS demonstrates the fundamentals transmission zeros and two poles. The transmission zero between the poles can be easily obtained from the ECM as:

$$f_0 \approx \frac{1}{2\pi\sqrt{L_s C_s}} \quad (2)$$

At the upper passband, the serie $L_o$, $C_s$ is viewed as opened circuit whereas the lower frequency is due to the interaction between both resonators. As a result:

$$f_u \approx \frac{1}{2\pi\sqrt{L_p C_p}} \quad (3)$$

$$f_l \approx \frac{1}{2\pi\sqrt{(L_p + L_s)C_s}} \quad (4)$$

The value of the electrical parameters $L_s$ and $C_s$ can be approximated using equations derived for subwavelength element found in [14]:

$$L_s = \frac{a}{2\pi}\mu_o\mu_{reff}\ln\left(\csc\left(\frac{\pi s}{2a}\right)\right) \quad (5)$$

where $\mu_o$ is the free space permeability and $\mu_{reff}$ is the effective permeability of the dielectric material which in this case is unity,

$$C_s = \frac{2\omega}{\pi}\varepsilon_o\varepsilon_{reff}\ln\left(\csc\left(\frac{\pi g}{2a}\right)\right) \quad (6)$$

$$\varepsilon_{reff} = \frac{\varepsilon_r + 1}{2} \quad (7)$$

where $\varepsilon_o$ is the free space permittivity and $\varepsilon_{reff}$ is the effective dielectric constant of the substrate. Then the constitute element of the bandpass resonator are given by [15]:

$$L_p = \frac{a}{2\pi}\mu_o\mu_{reff}\ln\left(\csc\left(\frac{\pi g}{2a}\right)\right) \quad (8)$$

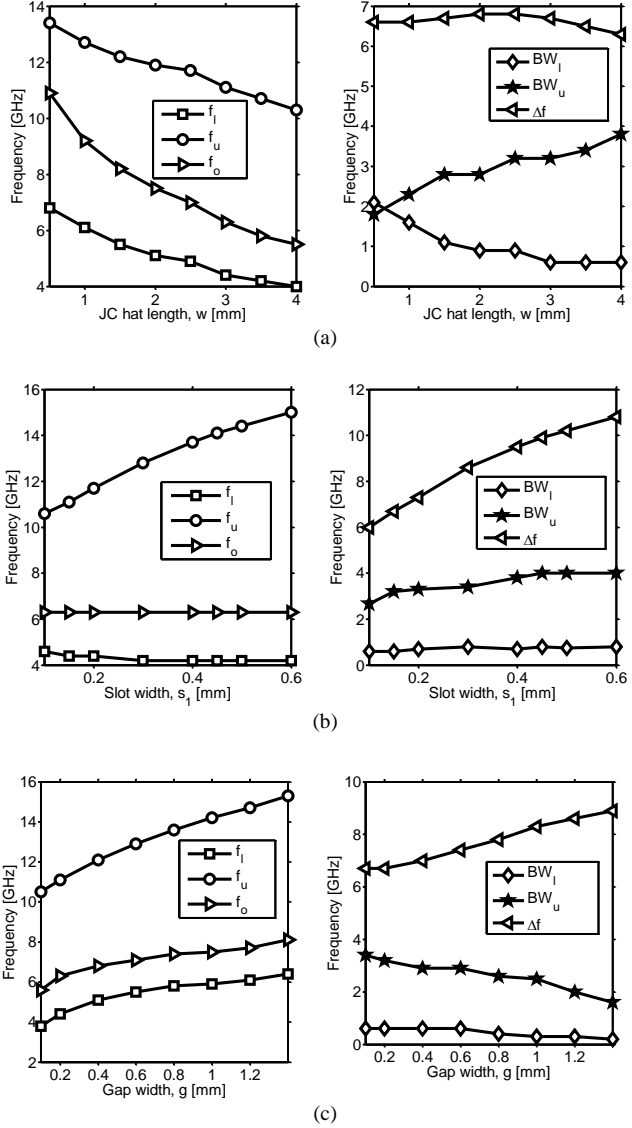

Fig. 3. Effect of the parametric on the frequency response of the FSS. (a) JC hat length $\omega$. (b) Slot width $s_1$ of the top layer. (c) Gap width g between 2 adjacent JC.

$$C_p = \frac{(a - g - s_1)}{\pi} \varepsilon_o \varepsilon_{reff} \ln\left(\csc\left(\frac{\pi s_1}{a - g - s_1}\right)\right) \qquad (9)$$

This analysis provides qualitative insight of the filter's characteristics. The lower band is mostly affected by the value of $L_p$ and $C_p$. The change of $C_p$ value only affects the center frequency of the upper band of operation. Also by increasing the value of $L_p$ both center frequencies decrease but the location of the transmission zero between both bands does not change. This transmission zero is only affected by the value of $L_s$ and $C_s$. Since each electrical parameter can be mapped to physical parameters by employing (5) - (9), this analysis is supported by a more accurate parametric study based on full wave simulation using HFSS as will be conveyed in the next section.

## III. PARAMETRIZATION RESULTS

A detailed parametric study correlating the frequency response behavior to the proposed FSS's elemental structures is investigated. Characteristics of the frequency response included center frequencies at $f_l$, $f_u$, and their bandwidth $BW_l$, $BW_u$ respectively, location of the transmission null between the two bands $f_o$ and the band separation $\Delta f$ between both center frequencies. The nominal dimension of the unit cell is set to be: $a = 5$ mm, $w = 3$ mm, $s = 0.15$ mm, $s_1 = 0.15$ mm, $g = 0.2$ mm, and substrate with thickness $h = 0.254$ mm and dielectric constant $\varepsilon_r = 10.2$.

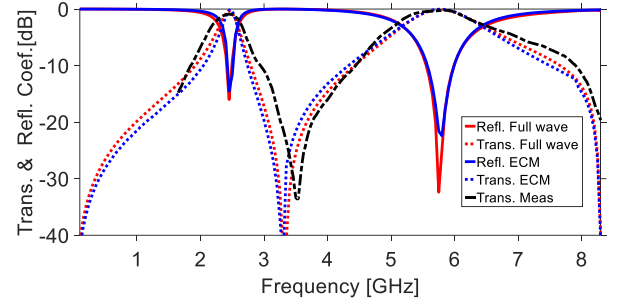

Fig. 4. Transmission and reflection coefficient of the FSS showing full wave and ECM simulation result, and measured transmission coefficient result for TEM plane wave. $Ls = 4.9$ nH, $Cs = 0.5$ pF, $Lp = 4$ nH, $Cp = 0.35$ pF; $L = 0.8$ nH; Physical parameters: $a = 8.5$ mm, $w = 6.8$ mm, $s = 0.3$ mm, $s_1 = 0.2$ mm, $g = 0.5$ mm, h = 0.635 mm, $\varepsilon_r = 10.2$.

### A. Effect of the JC hat length ($\omega$)

Fig. 3(a) shows the effect of the variation of the JC hat length $\omega$ from 0.5 mm to 4 mm on the frequency response characteristics of the design. The center frequencies $f_l$ and $f_u$ and the frequency of the transmission zero $f_o$, all decrease gradually thus reduce the element electrical size of the design. The spectral gap between the dual bands remains the same around 6.5 GHz. While the bandwidth at the upper band increases over a range of 2:1, the bandwidth at the lower band decreases at the same rate.

### B. Effect of the slot width $s_1$

Fig. 3(b) shows that only the center frequency at the upper band is affected by increasing $s_1$ from 0.1mm to 0.6mm. A significant increase of $f_u$ from 10.6 GHz to 15 GHz as well an increase in the band separation between both center frequencies. All the other factors remain virtually unchanged. This approach completely decouples both bands of operation which is consistent with the ECM analysis. The value of $C_p$ mainly depends on the separation gap $s_1$.

### C. Effect of the gap width between 2 adjacent JC (g)

Fig. 3(c) shows that both center frequencies increase with the gap width between the neighboring JC. The effect is more pronounced at the upper band. This effect is foreseen since (6) and (8) indicate a decrease of the value of $L_p$ and $C_s$ as the value of $g$ increases.

The outcomes from this investigation shows that design flexibility is achieved by logically altering the physical parameters. Thus a dual band FSS can be easily designed to meet most of the design specifications as the structure by means of its physical lengths allows controllability of the poles and zeros of its transfer function.

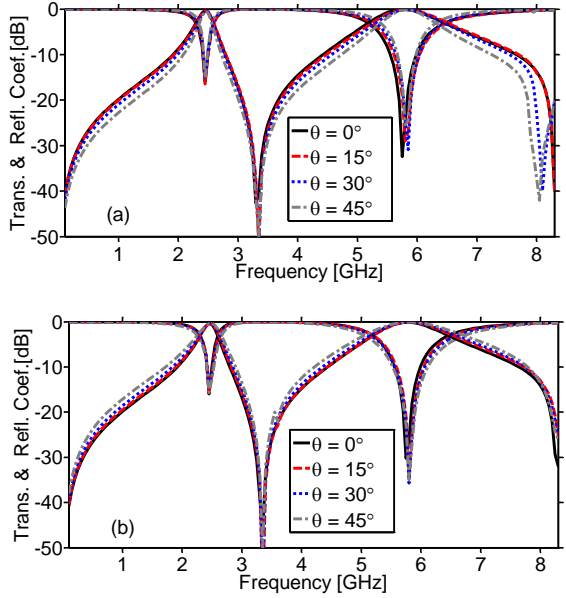

Fig. 5. Simulated reflection and transmission coefficients of the first order MEFSS as a function of incidence angle. (a) TE polarization, (b) TM polarization.

## IV. NUMERICAL SIMULATION AND MEASUREMENT RESULTS

### A. Implementation and simulation results of the prototype

Insight of the filter's characteristics obtained in previous section is applied to the design of a dual band MEFSS operating at $S$ and $C$ band with center frequencies $f_l = 2.4$ GHz and $f_u = 5.8$ GHz respectively. The band of operations of the FSS is chosen accordingly as it is used for many satellites communications transmissions, radar and WLAN applications. The approximated design dimensions are obtained using (5) - (9) and then a quick simple tuning was needed for the final dimensions. The substrate used is a 0.635 mm thick R06010 from Rogers Corp. laminated with 0.5 oz copper. This material has a relative permittivity value of $\varepsilon_r = 10.2$ and loss tangent of $tan\delta = 0.0023$. The periodicity of the unit cell is about $\lambda_l/15$ and the overall design thickness is in the order of $\lambda_l/200$ where $\lambda_l$ is the free space wavelength at $f_l = 2.4$ GHz. Fig. 4 shows the full wave simulation result of the filter performance compared with its ECM simulation result using a circuit simulator Agilent's Advanced Design Systems (ADS). There is strong consistency between ECM results and full wave simulations results. Results predicts a dual band performance operating at $S$ and $C$ band with center frequencies at $f_l = 2.4$ GHz and $f_u = 5.8$ GHz. The insertion loss at both resonances is 0.3 dB. The 3 dB percent bandwidth are 17% at lower band and 32% at upper band. The performance of the design is also investigated across different angles of incidence for both transverse electric (TE) and transverse magnetic (TM) polarizations as shown in Fig. 5. Simulations predicts good filter response for both TE and TM polarizations when the spatial filter is illuminated from various oblique angles ($0^\circ \leq \theta \leq 45^\circ$). This result is expected due to the electrically small unit cell dimensions and periodicity.

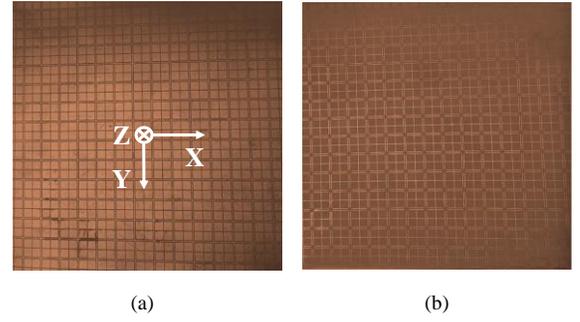

Fig. 6. Fabricated sample. (a) First layer, (b) Second layer.

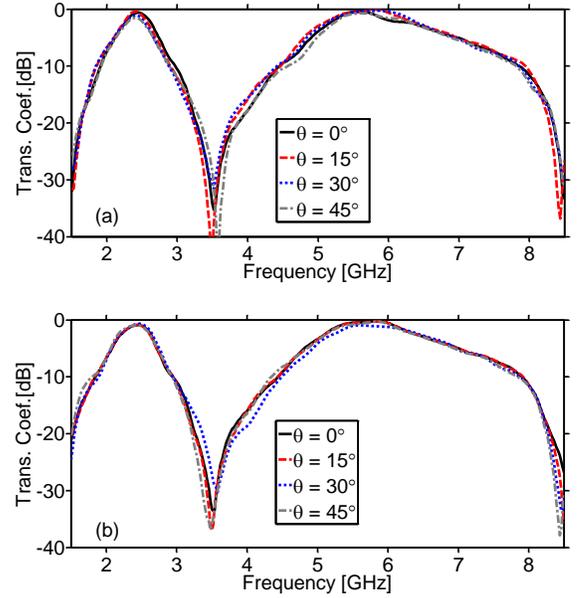

Fig. 7. Measured transmission coefficients of the first order MEFSS as a function of incidence angle. (a) TE polarization, (b) TM polarization.

### B. Fabrication and measurement results

To validate the proposed design, an array of 15 x 15 elements is fabricated on 5 inch x 5 inch dielectric (R06010) and measured in free space environment. The front and back layer of the sample are shown in Fig. 6. The prototype fits perfectly inside a 5 x 5 inch center aperture carved on a large wood fixture. Large wood fixture serves as FSS fixture and in minimizing the direct coupling between the Tx and Rx horn antennas. The front and back of the fixture is cover with 0.005 inch thick copper sheet. A pair of AEL H-1479 horn antennas operating from 1.5 GHz to 8.5 GHz were spaced 1.5 meters from each side of the fixture and are used to illuminate the sample with plane waves. The transmission scatter is measured using R&S ZVB20 network analyzer accurately calibrated within the range of operation of the horn antennas. The transmission coefficient is normalized with data obtained without the device under test. In order to eliminate ripples the normalized response is smoothed over 0.1 GHz. As can be observed, a good agreement between the measured, simulated and ECM results for TEM wave is observed. The measured results for oblique angle of incidence of the EM waves for both TE and TM polarizations are also shown in Fig. 7. A very stable dual band-pass spatial filter is achieved across scan angle with center frequencies located at $f_l = 2.45$ GHz and $f_u = 5.8$ GHz respectively. Measured insertion loss is 0.6 dB (slightly higher than the simulated insertion loss) at both bands with a 3 dB bandwidth of 18.5% at lower band and 32% at upper band.

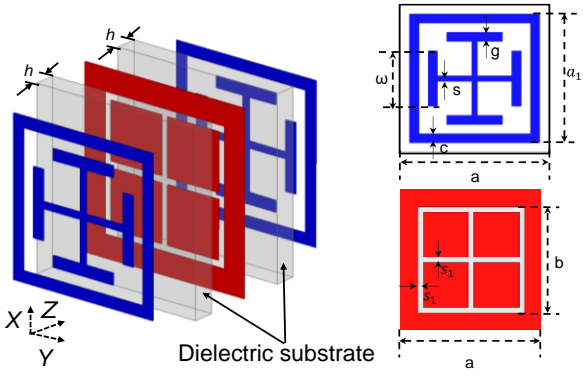

Fig. 8.   Topology of the proposed second order miniaturized element frequency selective surface along with the detailed geometry of the unit cells of each layer.

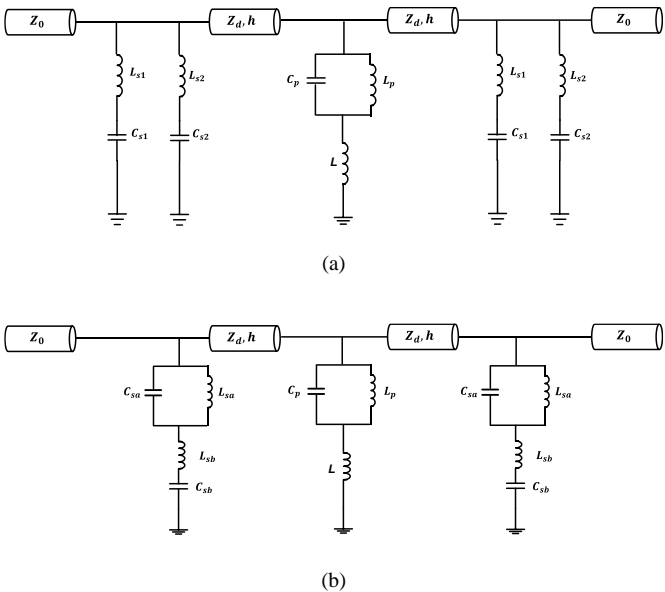

Fig. 9.   (a)Equivalent circuit model of the second order FSS proposed consisting of a 3 layers (1 bandpass resonator as a middle layer and 2 outer bandstop resonators). (b) The outer layers equivalent circuit model are converted to a hybrid resonator network composed of $L_{sa}$, $C_{sa}$ tank and serie $L_{sb}$, $C_{sb}$.

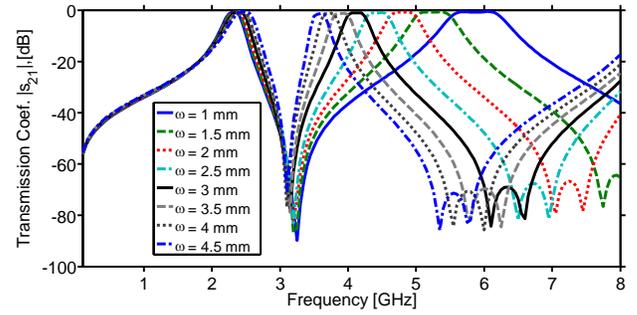

Fig. 10.   Frequency response of the second order MEFSS for TEM plane wave. Physical parameters of the unit cells are: $a = 7$ mm; $b = 4.4$ mm, $s_l = 0.2$ mm, $a_l = 6.8$ mm, $s = 0.2$ mm, $c = 0.2$ mm, $g = 0.5$ mm. Dielectric material: 0.5 oz. Rogers RT/duroid 6010, $h = 2.5$ mm, $\varepsilon_r = 10.2$. The effect of varying the length of the JC hat $\omega$ is observed only at the second band of operation.

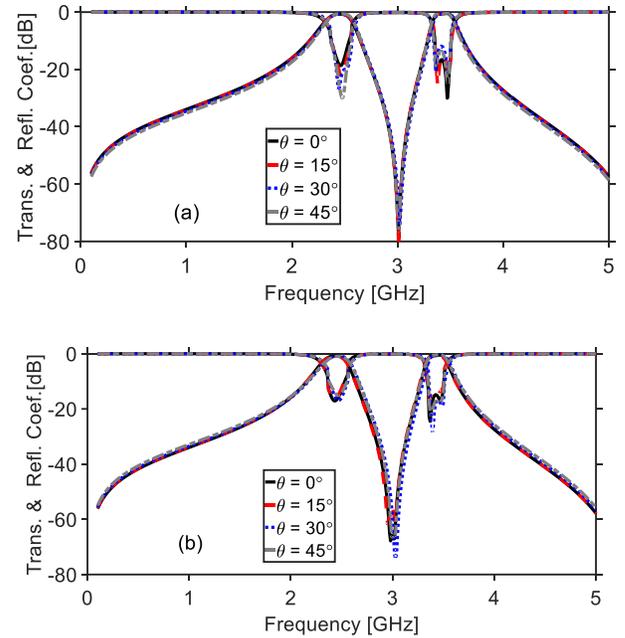

Fig. 11.   Simulated reflection and transmission coefficients of the second order dual band FSS (case of $\omega = 4.5$ mm) as a function of incidence angle. (a) TE polarization, (b) TM polarization.

## V. FILTER SELECTIVITY ENHANCEMENT TECHNIQUE

While the proposed design may be good candidate for systems requiring multiple operations, however for practical applications, selectivity at the upper band might present some challenges. Higher order filters response are preferred for application requiring sharper band skirt, and higher out-of-band suppression. Traditional methods cascade multiple FSS layers with quarter wavelength between each layer in order to achieve higher order responses. This method may be useful for application that operate in high frequencies. But at low frequency, this approach becomes problematic as it results in more weight and thickness of the structure. The latter approach leads to a deterioration of the filter performance for wave impinging at oblique incidence as it was the case in [9]. In this section, simple approach in obtaining second-order response at the upper band is implemented for higher selectivity of the band without any drastic change in the design size. This enhancement technique is achieved by loading the bandstop layer of the previous design with a square loop in a parallel combination. The modified layer is then alternated with the pass band layer resulting in 3 layers architecture as shown in Fig. 8. To better understand the working principle of this new design, a simple equivalent circuit model is derived in Fig. 9(a). For simplicity, the losses due to the conductor and dielectrics are omitted. The outer resonators can be modeled as two cascaded series $L_s$, $C_s$ [16]. Hereafter an additional attenuation pole has been generated in the upper stopband as shown in Fig.10. The second order transmission window at the upper band can also be explained by transforming the ECM in Fig. 9(a) to the equivalent network shown in Fig. 9(b) [12]. In this network the outer layers are modeled as a hybrid composed of parallel resonator (comprised of $L_{sa}$, $C_{sa}$) and series resonator (comprised of $L_{sb}$, $C_{sb}$). As a result a typical third order bandpass resonator filter is obtained. However since the separation between adjacent layers is much less than a quarter wavelength, the three transmission poles are located far apart from each other. In this case they are deliberately designed to be divided in two groups comprised of one pole in the first band and two poles in the second band. Thereby, a dual bandpass response is obtained. Due to the lack of comprehensive dual bandpass filter synthesis theory from the equivalent circuit obtained in Fig. 9, a circuit simulator Agilent's Advanced Design Systems in conjunction with full wave EM simulation are iteratively used. The physicals dimensions are obtained using (5) - (9) and approximated analytical expressions of the hybrid constitute elements available in [16]. As predicted, the full wave EM simulation results including dielectric and ohmic loss, shows higher selectivity at the upper band due to the additional transmission and attenuation pole. In addition, agility to control individually the operation bands is preserved. First, the dimensions of the design elements are optimized to locate the first band of operation centered at 2.4 GHz. Then by simply sweeping the length of the JC hat $\omega$ from 1 mm to 4.5 mm, the second band can be tuned from 6 GHz to 3.5 GHz enabling a very close or moderate spaced band of operation $\Delta f$. Besides, the first band of operation and the transmission zero between both bands are not affected by the variation of $\omega$ as can be seen in Fig. 10. The maximum insertion loss at the operation bands is about 0.9 dB. The 3 dB percent bandwidth is 10 % at the lower band and vary from 7-15% at the upper band during the tuning process. The scan of the incident angle ($0° \leq \theta \leq 45°$) is performed for the case where the spectral gap between both bands is about 1 GHz ($\omega$=4.5mm). Simulations predicts good filter response ($f_l$ = 2.45 GHz and $f_u$ = 3.4 GHz) for both TE and TM polarizations (Fig 11). This low profile design has a total thickness of 5 mm (equivalent to electrical thickness of $\lambda_l$/25) and the periodicity in the order of $\lambda_l$/18 where $\lambda_l$ is the free space wavelength at 2.4 GHz.

## VI. Conclusion

In this communication, a dual-band spatial filter technique based on modified Complementary Frequency Selective Surface (CFSS) is tailored to design a low-profile dual-band spatial filter operating at $S$ and $C$ bands. This novel design offers control over the location of both center frequencies as well as the poles of its transfer function. Due to the very small size of a unit cell compared to the wavelength, a quasi-static ECM is used to predict the behavior of the designed FSS. The proposed FSS delivers very stable filter performance for waves impinging from oblique angles while enabling very close or significant spectral gap between the operation bands. A low profile design with enhanced performance is also proposed. This structure exhibits a superior frequency response in term of selectivity of the bands and higher out of band rejection with the ability to independently control both bands of operation as well. A very sharp transmission zero with a very closely spaced band separation ($\Delta f$=1 GHz) can be achieved. The performance of these categories of spatial filter make them good candidates for radar and WLAN applications requiring compact structure, low profile and multiple bands of operations.